\documentstyle[11pt,iopconf]{article}

\newcommand{\beq}{\begin{equation}}
\newcommand{\beqar}{\begin{eqnarray}}
\newcommand{\eeq}[1]{\label{#1}\end{equation}}
\newcommand{\eeqar}[1]{\label{#1} \end{eqnarray}}

\newcommand{\oript}{$\top$}
\newcommand{\oripp}{$^|_|$}
\newcommand{\oriet}{$\dashv$}
\newcommand{\oriep}{$||$}
\newcommand{\oriec}{
  \hspace{0.10ex}\makebox[0in][c]{$-$}\makebox[0in][c]{$|$}\hspace{0.90ex}}
\newcommand{\orip}{$\stackrel{\circ}{_|}$}
\newcommand{\orie}{$\circ \! |$}
\newcommand{\oris}{$\circ \! \bigcirc$}

\begin{document}
\renewcommand{\topfraction}{0.95}
\renewcommand{\textfraction}{0.05}
\pagestyle{myheadings}
\thispagestyle{empty}
\markboth
{\protect\small\it A. Iwamoto, P. M\"{o}ller, J. R. Nix, and H. Sagawa /
Collisions of deformed nuclei}
{\protect\small\it A. Iwamoto, P. M\"{o}ller, J. R. Nix, and H. Sagawa /
Collisions of deformed nuclei}
\title{Collisions of Deformed
Nuclei:\\ A Path to the Far Side of the Superheavy Island}
\author{
Akira Iwamoto\dag, Peter M\"{o}ller\dag\ddag\S\P,
J. Rayford Nix\P, and Hiroyuki Sagawa\ddag}
\affil{\dag Advanced Science Research Center,
Japan Atomic Energy Research Institute, Tokai,
Naka-gun, Ibaraki, 319-11 Japan}
\affil{\ddag Center for Mathematical Sciences, University of Aizu,
Aizu-Wakamatsu, Fukushima 965-80, Japan}
\affil{\S P. Moller Scientific Computing and Graphics, Inc.,
P. O. Box 1440, Los Alamos, NM 87544, USA}
\affil{\P Theoretical Division,
Los Alamos National Laboratory, Los Alamos, NM 87545, USA}
\beginabstract
A detailed understanding of complete fusion cross sections
in heavy-ion  collisions requires a consideration of the
effects of the deformation of the projectile and target.
Our aim here is to show that deformation
and orientation of the colliding nuclei have a very significant
effect on the fusion-barrier height and on the compactness of
the touching configuration.
To facilitate discussions of fusion configurations of deformed nuclei,
we develop a classification scheme
and introduce a notation convention for these configurations.
We discuss particular
deformations and orientations that lead to
compact touching configurations and to
fusion-barrier heights that correspond to fairly low excitation energies
of the compound systems. Such configurations should be the most
favorable for producing superheavy elements. We analyze a few
projectile-target combinations whose deformations allow favorable
entrance-channel configurations and whose proton and neutron numbers
lead to  compound systems in a  part of the superheavy region
where $\alpha$ half-lives are calculated to be
observable, that is, longer than 1 $\mu$s.

\endabstract

\section{Introduction}
The last five elements that   have been
discovered [1--5]
were all formed in cold-fusion reactions between spherical nuclei.
As the proton number increases, the cross section for heavy-element
production decreases. For example, element 107 was produced with a 167 pb
cross section [1],
whereas for element 111 the production cross section
was only 2--3 pb [5].
There is reason to suspect that few additional new elements can
be reached in reactions between spherical nuclei because of the strong
decreasing trend of the cross sections.

In a fusion reaction several processes influence
 the evaporation-residue cross section $\sigma_{\rm er}$ for
heavy-element formation. One may write
\beqar
\sigma_{\rm cn} = \sigma_{\rm fus}(1-P_{\rm ff})  \nonumber \\
\sigma_{\rm er} = \sigma_{\rm cn}(1-P_{\rm cf})
\eeqar{crossform}
Here $\sigma_{\rm cn}$ is the cross section for forming a compound nucleus,
$\sigma_{\rm fus}$ is the fusion cross
section corresponding to the  coalescence
of the target and projectile,
$P_{\rm ff}$ is the probability for fast fission
prior to forming a compound system, and
$P_{\rm cf}$ is the probability for fission during the deexcitation
of the compound nucleus through neutron and gamma emission.

In fusion reactions where the number of protons in the projectile and
target add up to  about 100,  the overwhelming inelastic cross-section
component is fusion-fission,
which includes both processes in which
the coalesced system undergoes fast
fission prior to compound-nucleus formation and the compound nucleus
undergoes fission before complete deexcitation.
For bombarding energies close to the
Coulomb barrier $\sigma_{\rm fus}$ is in the millibarn range, whereas
$\sigma_{\rm er}$ is in the picobarn range.
Because small changes in the fusion-reaction parameters and the
nuclear-structure parameters may  change $P_{\rm ff}$ and $P_{\rm cf}$
by orders of magnitude but leave $\sigma_{\rm fus}$
essentially unchanged when the
incident energy measured from the barrier top is the same,
$\sigma_{\rm fus}$ is relatively unimportant in heavy-element
formation.  Instead, it is crucial to minimize $P_{\rm ff}$ and $P_{\rm
cf}$, which are very close to 1 in reactions leading to compound systems
with more than 200 nucleons.

In general, large negative shell
corrections of the ground state of the compound system decrease $P_{\rm
cf}$ and consequently enhance the heavy-element-formation cross section
$\sigma_{\rm er}$. The compound-nucleus fission cross section
is usually modeled in terms of a level-density formalism.

Models for $P_{\rm ff}$ are much less reliable than models for $P_{\rm
cf}$.  In a classical picture a necessary condition for complete fusion
and the formation of a compound nucleus is that the fusing system
evolves into a configuration inside the fission saddle point in a
multi-dimensional deformation space
[6--9].  In heavy-ion collisions
where the projectile and target are of roughly equal size  and with a
nucleon number $A$ above about 100, the touching configuration lies on
the side of a steep hill outside the fission saddle point
[10].  For energies just above the Coulomb barrier this
topographical feature results in a trajectory that is deflected away
from the direction that would lead to the spherical shape. Instead, it
proceeds from the touching configuration to the fission valley, so that
no compound-nucleus formation occurs.

There are two simple possibilities for overcoming the above limitation
to compound-nucleus formation and increasing the cross section for
heavy-element production. First, if the projectile energy is increased
sufficiently high, the trajectory will in the absence of dissipative
forces pass inside the fission saddle point. However, dissipation may
make such trajectories difficult to realize. Second, highly asymmetric
touching configurations may be sufficiently close to the ground-state
shape of the compound nucleus that the touching configuration is inside
the fission saddle point. Thus, these two simple principles would
suggest that to produce elements in the superheavy region one should
select highly asymmetric configurations and increase the projectile
energy above the Coulomb barrier. However, high excitation energies and
the resulting high angular momentum of the compound system may favor
fission instead of de-excitation by neutron emission. In the
cold-fusion approach that led to the identification of the five
heaviest elements, the very nature of cold fusion leads to a low
excitation energy of the compound system.  The entrance-channel
configuration is also fairly asymmetric and compact. However, the
maximum cross section for the production of the heaviest elements
occurs at sub-barrier energies as very rare, non-classical events.

In our study here we will mainly argue that in the fusion of deformed
nuclei there are particular deformations and relative orientations of
the target and projectile that favor compound-nucleus formation, that
is, decrease $P_{\rm ff}$.  Our discussion above revealed that from
very general principles one can expect that heavy-element production in
heavy-ion reactions is most favorable when the touching configuration
is compact.  The excitation energy of the compound system should be
high enough to allow a trajectory inside the fission saddle point, but
otherwise as low as possible to reduce the fission branch of the
compound system. A spherical picture of nuclei in heavy-ion collisions
allows few new possibilities for very-heavy-element production beyond
what has already been accomplished. It is therefore of interest to
investigate if consideration of deformation will identify
entrance-channel configurations that have some possibility of being
more favorable for heavy-element production than is expected from the
spherical picture.  To facilitate the discussion of deformed fusion
configurations we introduce a classification scheme, notation and
terminology.

\section{Fusion configurations of
  deformed nuclei:\protect\\
Classification, notation and terminology}

Obviously, the multi-dimensional fusion potential is a continuous
function of the incident direction and orientation of the
projectile nucleus and of the deformation of the
projectile and target. However, to allow the identification and discussion
of major physical effects it is useful to identify and
study a few limiting
situations.

\subsection{Limiting fusion configurations}

Our discussion of specific cases below will show
that for prolate shapes there are significant differences
in the fusion process depending on the sign of the hexadecapole
moment.
Nuclei with a
large negative hexadecapole moment
develop a neck which allows a close approach.
As a result the fusion configuration
for some orientations of the projectile-target combinations
is considerably more compact than the corresponding
configurations for shapes with large positive
hexadecapole moments.
Thus, we identify four limiting situations as far as deformations
are concerned. They are:
\begin{enumerate}
\item
Well-developed oblate shapes
\item
Spherical shape
\item
Well-developed prolate
shapes with large negative
hexadecapole moments $Q_4$
\item
Well-developed prolate
shapes with large  positive
hexadecapole moments $Q_4$
\end{enumerate}
Furthermore, we assume mass symmetry
and axial symmetry as this is consistent with the vast majority
of nuclear ground-state configurations.

In our studies here we use alternatively the
Nilsson perturbed-spheroid parameterization
$\epsilon$ [11] and the $\beta$  parameterization
 to  generate deformed nuclear shapes.
In the $\beta$ parameterization, assuming
axial symmetry, the radius vector
$R(\theta,\phi)$ to the nuclear surface is defined by
\beq
R(\theta,\phi) = R_0
\left[ 1 +\sum_{l=1}^\infty \beta_{l}Y_l^0(\theta,\phi) \right]
\eeq{radvec}
where $R_0$ is deformation dependent so as to conserve the
volume inside the nuclear surface. The variation in $R_0$ due to
volume conservation is only a fraction of one percent.
The definition of the $\epsilon$ parameterization is more complicated.
A  recent, extensive presentation  is given in Ref.~[12].
One should note that large positive $Q_4$ corresponds to positive
$\beta_4$ but to negative $\epsilon_4$ and that
large negative $Q_4$ corresponds to negative
$\beta_4$ but to positive $\epsilon_4$.

As limiting orientations we consider only situations where
the projectile center is on the x, y or z axis of the target and
orientations of the projectile where the projectile symmetry axis is either
parallel to or perpendicular to the target symmetry axis.
Since we restrict ourselves to axial symmetry, configurations with
the projectile center  located on the x or y axis are identical.
If the projectile is located in the equatorial region of
the target it can be oriented in three major orientations,
and if it is located in the polar
region it can be oriented in two major orientations. Thus, for a particular
projectile-target deformation combination there are five
possible limiting configurations.

Because there are five orientations and three major types of deformations
for both projectile and target  there are 45 different
configurations when the projectile and target
are deformed and of unequal mass.
When the projectile and target are of equal
mass, one would at first sight expect 30 different configurations.
We later show that in the case of equal projectile and target mass
there are three pairs of configurations where the two configurations
in the pairs are identical. Therefore, there are in this case only 27 deformed
configurations that are different.
Situations where either the projectile or target
is deformed add another six configurations and, finally,
we designate a spherical target and a spherical projectile
as  a separate configuration. Thus, in our classification scheme
we find 34 configurations of projectile and target in
heavy-ion collisions that are different
also in the special case of equal projectile
and target mass.  For the case of unequal projectile and target mass
one may wish to count a total of 45 different deformed configurations,
for a total of 52 different fusion configurations.

We will in a separate study systematically
review the barrier parameters of
these  configurations for projectiles and targets throughout
the periodic system. Here, we will just discuss   a few
configurations with potential importance for very-heavy-element
production. However, to be able to simply and transparently refer
to any of the limiting configurations we start by introducing a notation
convention for deformed fusion configurations.

\subsection{Notation for deformed fusion configurations}

We denote a particular fusion configuration by
[P,T,O], where the three letters stand for
Projectile deformation, Target deformation, and relative
Orientation of the projectile-target combination.
For configurations where the projectile or target or both
are spherical, the number of different limiting orientations
is less than when both the projectile and target are deformed.
It is therefore most clear to introduce notation that distinguishes
between these possibilities.
The following values are
possible for the three entities P, T and O:
\begin{description}
\item[P and T]\mbox{ }\\
Oblate:\dotfill o\\
Spherical \dotfill s\\
Prolate with negative $Q_4$ \dotfill p$^-$\\
Prolate with positive $Q_4$ \dotfill p$^+$
\item[O] {\bf Spherical projectile and spherical target}\\
Spherical (s) \dotfill
    \oris

\item[O] {\bf Spherical-deformed projectile-target combination}\\
Polar (p) \dotfill
    \orip
\\
Equatorial (e) \dotfill
     \orie

\item[O] {\bf Deformed-deformed projectile-target configuration}\\
Polar-transverse (pt) \dotfill
    \oript \\
Polar-parallel (pp) \dotfill
     \oripp \\
Equatorial-transverse (et) \dotfill
            \oriet \\
Equatorial-parallel (ep) \dotfill
               \oriep   \\
Equatorial-cross (ec) \dotfill
                \oriec

\end{description}

We prefer the graphical short-hand notation given in the table above
for the different orientations, but we also provide in parenthesis
an alternative notation, based on letters only.

In Fig.~\ref{orisphe} we show the seven
different configurations that can occur with
a spherical projectile. We have sandwiched the familiar
spherical-projectile spherical-target case between the prolate-target and
oblate-target configurations in the top row so that the appearance
of the configurations evolves smoothly from the polar, spherical-prolate
positive-hexadecapole configuration [s,p$^-$,\orip]
on the extreme left to the polar, spherical-projectile oblate-target
configuration [s,o,\orip] on the far right.

In Figs.~\ref{oripppp}--\ref{orioo} we show the 45 different
configurations occurring in our scheme when both the projectile and target
are deformed and of unequal mass.
In the case of equal projectile and target mass
the configuration [p$^+$,p$^-$,\oript] and [p$^-$,p$^+$,\oriet], for
example, are
identical.
Indeed, in this case all the configurations
[p$^-$,p$^+$,any] have a corresponding configuration
[p$^+$,p$^-$,any], and other similar correspondences also occur.
Therefore, when the projectile and target mass are equal,
one need not consider any configuration that occurs in
Figs.~\ref{oripmpp}, \ref{oriopp} and \ref{oriopm} because equivalent
configurations will have occurred in an earlier figure.
For equal-mass projectile-target
combinations the configurations
[p$^+$,p$^+$,\oript],
[p$^-$,p$^-$,\oript] and
[o,o,\oript] are equivalent to
[p$^+$,p$^+$,\oriet],
[p$^-$,p$^-$,\oriet] and
[o,o,\oriet], respectively. This is the reason there are only
27 different configurations when the projectile and target are of
equal mass.

In Figs. \ref{orisphe}--\ref{orioo} we use the $\beta$ parameterization
to describe the nuclear shape. Volume conservation has not
been applied in these and subsequent figures of nuclear shapes,
but this is an
insignificant approximation since volume conservation only changes
$R_0$ by fractions of a percent
for the deformations considered. However, in energy
calculations it is essential to include volume conservation, as we do
in our calculations here.
As representative deformations we make the following choices.
As the prolate--positive hexadecapole deformation
p$^+$ we choose $\beta_2=0.30$ and $\beta_4=0.11$.
This corresponds to the experimentally determined deformation of
$^{154}$Sm [12]. The prolate-negative hexadecapole
deformation
p$^-$ is chosen as $\beta_2=0.24$ and $\beta_4=-0.09$, corresponding
to the experimentally determined deformation of $^{186}$W [12].
Finally, as a representative oblate deformation we have selected
$\beta_2=-0.25$ and $\beta_4=0.0$. The ratio between $R_0$ of the
projectile and target  is 0.7.

\section{Deformation and heavy-ion collisions}

Although the implications of deformation on cross sections for
superheavy-element production have not been very extensively
considered so far, deformation certainly is already known to affect
fusion cross sections leading to somewhat lighter compound systems.
For example, a clear signature of the importance of deformation
effects in heavy-ion reactions is the enhancement of sub-barrier
fusion cross sections, for which deformation often plays a major role.
It may be useful to observe that the designation sub-barrier is
somewhat of a misnomer.  An implicit assumption behind this
designation is that both projectile and target nuclei are spherical.
Furthermore, if the measured cross section at energies below the
maximum of this assumed spherical fusion barrier is higher than the calculated
cross section for this configuration then the term {\it enhanced
sub-barrier fusion} is used.  In a more realistic picture one can in
many cases show that (1) the energy is not sub-barrier and (2) the
measured cross section is not enhanced.  To illustrate these features
we select the reaction $^{16}{\rm O}+\mbox{}^{154}$Sm.

\subsection{Deformation and the fusion potential-energy surface}

We show in Fig.~\ref{potsmod} a calculated fusion potential-energy
surface for the reaction $^{16}{\rm O}+\mbox{}^{154}$Sm.  The
potential-energy surface is calculated in our previously developed
model [12] for the macroscopic potential energy between
two arbitrarily oriented, deformed heavy ions. The potential shown in
Fig.~\ref{potsmod} is considerably simpler than the most general case,
since the projectile is spherical and the target has both mass and
axial symmetry. Thus, for this system the fusion potential is
completely specified by this two-dimensional figure.

We present in Table 1 four fusion-barrier quantities for particular
orientations between the projectile and target. Each line corresponds to
\begin{table}[t]
{\small
\begin{center}
\caption[taberr]{\baselineskip=12pt\small
Comparison of entrance-channel fusion configurations.
When the sign $<$ is given in the column for $R_{\rm max}$
and $>$ is given in the column for $V_{\rm max}$
it means that the maximum of the fusion barrier occurs inside
the touching point and consequently is higher than
the potential of the touching configuration.
\\}
\begin{tabular}{rrrrrrrrrcrrrrr}
\hline\\[-0.07in]
\multicolumn{4}{c}{Target}      & &
\multicolumn{5}{c}{Projectile} & &
\multicolumn{4}{c}{Barrier}
\\[0.08in]
\cline{1-4} \cline{6-10} \cline{12-15}\\[-0.07in]
             &
\multicolumn{1}{c}{$\epsilon_2$} &
\multicolumn{1}{c}{$\epsilon_4$} &
\multicolumn{1}{c}{$\epsilon_6$} &
             &
             &
\multicolumn{1}{c}{$\epsilon_2$} &
\multicolumn{1}{c}{$\epsilon_4$} &
\multicolumn{1}{c}{$\epsilon_6$} &
\multicolumn{1}{c}{Or.}&
             &
\multicolumn{1}{c}{$R_{\rm max}$}&
\multicolumn{1}{c}{$V_{\rm max}$}&
\multicolumn{1}{c}{$R_{\rm t}$}&
\multicolumn{1}{c}{$V_{\rm t}$}
\\
  &
  &
  &
  &
  &
  &
  &
  &
  &
\multicolumn{1}{c}{ }  &
  &
\multicolumn{1}{c}{(fm)}  &
\multicolumn{1}{c}{(MeV)} &
\multicolumn{1}{c}{(fm)}  &
\multicolumn{1}{c}{(MeV)}
\\[0.08in]
\hline\\[-0.07in]
 $^{154}$Sm &
0.000       &
0.000       &
0.000       &
           &
 $^{16}$O &
0.000       &
0.000       &
0.000       &
\oris       &
           &
$10.54$       &
$62.21$       &
9.14     &
56.22       \\
%
%
 $^{154}$Sm &
0.250       &
$-0.067$       &
0.030       &
           &
 $^{16}$O &
0.000       &
0.000       &
0.000       &
\orie       &
           &
$10.10$       &
$63.29$       &
8.80     &
57.90       \\
%
%
 $^{154}$Sm &
0.250       &
$-0.067$       &
0.030       &
           &
 $^{16}$O &
0.000       &
0.000       &
0.000       &
{\footnotesize \orip}       &
           &
$11.87$       &
$57.18$       &
10.67     &
53.34       \\
%
%
 $^{150}$Nd &
0.000       &
0.000       &
0.000       &
           &
 $^{150}$Nd &
0.000       &
0.000       &
0.000       &
\oris       &
           &
$<$       &
$>$       &
12.33      &
379.10       \\
%
%
 $^{150}$Nd &
0.225       &
$-0.067$       &
0.025       &
           &
 $^{150}$Nd &
0.225       &
$-0.067$       &
0.025       &
{\footnotesize
\oriec }       &
           &
$<$          &
$>$          &
11.74      &
390.96
\\[1ex]
%
%
 $^{150}$Nd &
0.225          &
$0.200$        &
$-0.100$       &
               &
 $^{150}$Nd    &
0.225          &
$0.200$        &
$-0.100$       &
{
\oriec }  &
           &
11.69      &
399.51     &
10.29      &
383.98
\\
%
%
 $^{150}$Nd &
0.225          &
$0.100$        &
$-0.044$       &
               &
 $^{150}$Nd    &
0.225          &
$0.100$        &
$-0.044$       &
{
\oriec  }   &
           &
11.66      &
396.73     &
10.86      &
392.38
\\
%
%
 $^{186}$W &
0.208          &
$0.100$        &
$-0.044$       &
               &
 $^{110}$Pd    &
0.200          &
$0.027$        &
$-0.013$       &
{
\oript }       &
           &
$<$      &
$>$     &
12.29     &
358.13
\\
%
%
 $^{186}$W &
0.208          &
$0.100$        &
$-0.044$       &
               &
 $^{110}$Pd    &
0.200          &
$0.027$        &
$-0.013$       &
{\footnotesize
\oripp }       &
           &
$<$     &
$>$     &
13.46     &
342.61
\\
%
%
 $^{186}$W &
0.208          &
$0.100$        &
$-0.044$       &
               &
 $^{110}$Pd    &
0.200          &
$0.027$        &
$-0.013$       &
{
\oriet }       &
           &
$<$      &
$>$     &
12.15     &
359.01
\\[1ex]
%
%
 $^{186}$W &
0.208          &
$0.100$        &
$-0.044$       &
               &
 $^{110}$Pd    &
0.200          &
$0.027$        &
$-0.013$       &
{\footnotesize
\oriep }       &
           &
11.69      &
375.12     &
10.99     &
372.84
\\
%
%
 $^{186}$W &
0.208          &
$0.100$        &
$-0.044$       &
               &
 $^{110}$Pd    &
0.200          &
$0.027$        &
$-0.013$       &
{\footnotesize
\oriec }       &
           &
11.69      &
376.20     &
10.99     &
374.14
\\
%
%
 $^{186}$W &
0.000          &
$0.000$        &
$0.000$       &
               &
 $^{110}$Pd    &
0.000          &
$0.000$        &
$0.000$       &
\oris      &
           &
$<$      &
$>$     &
12.18     &
361.10
\\
%
%
 $^{192}$Os&
0.142          &
$0.073$        &
$-0.032$       &
               &
 $^{104}$Ru    &
0.233          &
$-0.013$        &
$0.012$       &
\oriec      &
           &
$11.72$      &
$367.82$     &
11.22     &
367.11
\\
%
%
 $^{186}$W &
0.208          &
$0.100$        &
$-0.044$       &
               &
 $^{104}$Ru    &
0.233          &
$-0.013$        &
$0.012$       &
{\footnotesize
\oriec }       &
           &
11.54      &
362.33     &
10.94     &
360.95
\\[1ex]
%
%
 $^{186}$W &
0.208          &
$0.100$        &
$-0.044$       &
               &
 $^{116}$Cd    &
$-0.233$          &
$ 0.053$        &
$-0.002$       &
{\footnotesize \oripp}       &
           &
12.44      &
379.14     &
11.84     &
377.30
\\
%
%
 $^{186}$W &
0.208          &
$0.100$        &
$-0.044$       &
               &
 $^{116}$Cd    &
$-0.233$          &
$ 0.053$        &
$-0.002$       &
\oriet      &
           &
11.44      &
397.59     &
10.54   &
391.47
\\
%
%
 $^{186}$W &
0.000          &
$0.000$        &
$0.000$       &
               &
 $^{116}$Cd    &
$0.000$          &
$ 0.000$        &
$0.000$       &
\oris       &
           &
$<$      &
$>$     &
12.28     &
375.27
\\
%
%
 $^{248}$Cm &
0.217          &
$-0.013$        &
$0.042$       &
               &
 $^{48}$Ca    &
$0.000$          &
$ 0.000$        &
$0.000$       &
{\footnotesize \orip }      &
           &
$13.31$      &
$198.33$     &
12.41     &
195.44
\\
%
%
 $^{248}$Cm &
0.217          &
$-0.013$        &
$0.042$       &
               &
 $^{48}$Ca    &
$0.000$          &
$ 0.000$        &
$0.000$       &
\orie      &
           &
$11.69$      &
$212.56$     &
11.09     &
211.23
\\[0.08in]
\hline
\end{tabular}\\[3ex]
\end{center}
}
\end{table}
one orientation and one incident direction.
The first eight columns specify the
projectile and target nuclei and the deformation used
for these nuclei in the calculation of
the fusion barrier.
The shapes of the projectile and target are given in the Nilsson
perturbed-spheroid parameterization [11].
The next column gives the relative orientation of
projectile and target
in the notation introduced above.
The last four
columns indicate (1) the distance between the centers-of-mass of the
projectile and target
at the maximum of the barrier, (2) the maximum of the fusion
barrier, (3) the center-of-mass distance when the projectile and target
just touch and (4) the fusion-barrier height at this point.

The first three lines of Table 1 show fusion-barrier data for the
reaction $^{16}{\rm O}+\mbox{}^{154}$Sm.  In the first line of the table we
show, for reference, the calculated barrier parameters for a
hypothetical spherical target shape. The second line gives the
fusion-barrier parameters for the configuration [s,p$^+$,\orie]
corresponding to the equatorial plane $z=0$ in Fig.~\ref{potsmod}
and the third line
corresponds to the potential in the  [s,p$^+$,\orip] configuration,
that is, to the
line $\rho=0$ in Fig.~\ref{potsmod}.

\subsection{Deformation and fusion cross sections}

In Fig.~\ref{crsmo} the measured and
calculated fusion cross sections corresponding to the fusion potential-energy
surface displayed in Fig.~\ref{potsmod} are presented.
Figure~\ref{crsmo} is taken from a
calculation [13] of fusion cross sections in reactions
of spherical projectiles and deformed targets.  It has no free
parameters except a simple translation in energy of the calculated
cross-section curves.
The cross section is obtained by integrating over angle the
transmission
coefficients which are determined by calculating the barrier
penetrability at each angular momentum by use of the WKB
approximation.
The deformed fusion potential is obtained in a
model calculation with no free parameters
and is the sum of the nuclear and Coulomb potentials according to
Ref.~[12] and
a centrifugal barrier term, which is treated
in the spherical limit.  The deformation parameters of
the target are taken from a recent calculation
of nuclear ground-state masses and deformations [14].
Obviously, there are large deformation effects both in the potential
energy and in the fusion cross section. Our model,
incorporating significant aspects of
deformation, accounts well for the ``enhancement'' of the
cross section relative to the fusion cross section
obtained for a hypothetical spherical target,
at least for energies down to the Coulomb barrier in
the polar direction.

However, our present model of fusion cross sections does not account
for fusion enhancement due to surface vibrations. For example, it cannot
account for the well-known enhancements in fusion cross sections
in such reactions as $^X{\rm Ni}\mbox{}+\mbox{}^Y$Ni of Ni isotopes, where
$X$ and $Y$ are close to 58. A recent study of such reactions is found in
Ref.~[15].

To further illustrate aspects of our unified approach to calculating
nuclear ground-state masses, shapes, reaction and decay $Q$ values,
$\beta$-decay rates, fission barriers, and fusion cross section
$\sigma_{\rm fus}$ we discuss some similarities and differences between
our fusion cross section calculation and a recent, detailed study
[16] of the reaction $^{16}{\rm O}+\mbox{}^{154}$Sm and
other reactions.  The latter study is based on a parameterized
distribution of fusion barriers, where deformation and several other
parameters are selected so as to obtain optimum agreement with data. In
our calculation the nuclear and Coulomb potentials are generated from
{\it calculated} nuclear ground-state shapes.  The parameters of our
potentials are identical to those of a global nuclear mass
calculation~[17,18]. The only free parameter in
our fusion cross-section calculation is a translation of the calculated
curve, corresponding to an adjustment of the calculated fusion-barrier
height.  In our calculation we take into account the three lowest even
multipole shape deformations and evaluate the integral expressions for
the Coulomb and nuclear interaction energies by numerical integration
to obtain the required accuracy.  In contrast, the study of
Ref.~[16] considers only $\beta_2$ and $\beta_4$ shape
parameters and the Coulomb interaction is evaluated only approximately
by expanding to order ${\beta_2}^2$ and $\beta_4$.

Despite some differences between our calculation and that of
Ref.~[16], the similarities are substantial. In particular,
we are in good agreement with their conclusions about the values of the
deformation parameters $\beta_2$ and $\beta_4$ for $^{154}$Sm and agree
with Ref.~[16] that earlier ambiguities about the values of
these deformation parameters as derived from fusion cross-section data
were largely due to inexact theoretical models for the fusion cross
sections. We obtain a difference of 6.56 MeV between the fusion-barrier
height in the equatorial and polar regions, whereas
Ref.~[16] obtains a difference of 8 MeV.

Many other calculations of the  fusion cross section for the reaction
$^{16}{\rm O}+\mbox{}^{154}$Sm have been published earlier. In most of these
studies a set of nuclear deformation parameters are determined by adjusting
the calculated curve to the measured cross-section values. Our purpose
in presenting our calculated cross section for this reaction here
is to show that our cross-section calculation based on {\it calculated}
deformation parameters agrees extremely well with the measured values.
Thus, our {\it calculated} nuclear shape of $^{154}$Sm should closely
resemble the actual shape. Because of this very good
agreement, our discussions below
of
the influence of higher multipole moments on $P_{\rm ff}$ can be expected
to be based on realistic ground-state deformations.
It has been shown earlier [19]
that the sign of $\beta_4$ significantly influences
the fusion cross section $\sigma_{\rm fus}$. Because
changing the sign  of $\beta_4$
under certain conditions may influence $P_{\rm ff}$
by several orders of magnitude but influences $\sigma_{\rm fus}$ much less,
we are below mainly concerned with how quadrupole and higher multipole
deformations affect $P_{\rm ff}$. One should note that increasing
$\epsilon_4$ while keeping $\epsilon_2$ fixed decreases the
fusion cross section, whereas, as we argue below, it can be expected to
increase compound-nucleus-formation probability or equivalently
decrease $P_{\rm ff}$. The effect of $\epsilon_4$ variations on $P_{\rm ff}$
is under certain circumstances
expected to be substantially larger than the (opposite) effect
on $\sigma_{\rm fus}$, so that there is an overall increase in
the evaporation-residue cross section $\sigma_{\rm er}$ for increasing
$\epsilon_4$.

\subsection{Gentle fusion?}

Because the evaporation-residue cross sections
in cold fusion between spherical projectiles and targets
drop so strongly
towards heavier nuclei, N{\"{o}}renberg [20,21]
suggested that ``gentle fusion''
of two well-deformed rare-earth nuclei in an equatorial-cross
orientation \oriec\
should be investigated because, he stated,
``this orientation leads to
the most compact touching configuration
out of all possible orientations of the two deformed nuclei.'' Consequently,
the
evaporation-residue cross sections may be sufficiently large to allow
detection.

We first observe that according to our calculations [14],
only the lightest nuclei in the rare-earth region would lead to
compound systems with $\alpha$ half-lives over 1 ${\mu}$s, which is
the approximate transit time from the target to detection area in the
SHIP experimental setup. Already the reaction $^{160}{\rm
  Gd}+\mbox{}^{160}$Gd$\rightarrow \mbox{}^{320 - x{\rm n}}128 + x{\rm
  n}$ leads to nuclei where the calculated
[14,22] $\alpha$-decay half-lives are less than
about 0.01 $\mu$s. To study the concept of gentle fusion we must
therefore select a reaction in the beginning of the rare-earth region,
so we choose the reaction $^{150}{\rm Nd}+\mbox{}^{150}$Nd to illustrate
N\"{o}renberg's suggestion.  We show the configuration of two
$^{150}$Nd nuclei  with calculated ground-state
shapes in Fig.~\ref{ndndgent}. The configuration is
[p$^+$,p$^+$,\oriec]
and is the one proposed by N{\"{o}}renberg as favorable
for SHE production. Calculated fusion-barrier data for the
hypothetical spherical case and the configuration in
Fig.~\ref{ndndgent} are found in Table 1, on lines 4 and 5,
respectively.

It is clear
that the fusion
configuration \oriec\
suggested by N{\"{o}}renberg is limited to
[p$^+$,p$^+$,\oriec] configurations, since
projectiles and targets must be chosen from the beginning of the
rare-earth region.
This configuration is not particularly compact relative to
a collision between similar-size spherical nuclei,
as is clear from  Figs. \ref{oripppp} and
\ref{ndndgent} and  Table 1.
Indeed, because of the large negative $\epsilon_4$ of the ground
state, which results in a bulging equatorial region
and a large positive hexadecapole moment, the configuration
in Fig.~\ref{ndndgent} is quite similar to the spherical configuration.
 This observation is supported by the
quantitative results in Table 1: the distance between mass centers of
the gentle fusion configuration is 11.74 fm, only 0.59 fm more compact
than the spherical configuration.

The idea that  configurations where deformed nuclei
touch each other in the equatorial regions are more compact than
some other configurations  and may therefore be
favorable for SHE production is not new. It was for instance mentioned
in Ref.~[12] in a discussion of the reaction
$^{48}{\rm Ca}+\mbox{}^{248}$Cm, and we will return to this reaction below.
Clearly, the fusion barrier for deformed systems
along a one-dimensional path will be very different
in the polar direction and in an
equatorial direction.
When the projectile is deformed the fusion barrier
will also depend strongly on the
orientation of the incident deformed projectile.

It is obvious that when colliding heavy ions have well-developed
prolate deformation, then the most compact configurations occur
when the point of touching is in the equatorial region of both
nuclei. Which relative orientation  of the two
nuclei, \oriec\ or {\oriep}, is the most
favorable is perhaps not known at present. However, the orientation
suggested by N\"{o}renberg is one possible favorable configuration,
but its properties will depend strongly on the value of the
hexadecapole deformation, that is, in our case on the value of the
deformation parameter $\epsilon_4$. Large negative values of
$\epsilon_4$ correspond to bulging equatorial regions, whereas
positive values lead to neck formation. We now look at the latter, more
compact configurations.

\subsection{Hugging fusion!}

To clearly illustrate the effect of large positive values of
the deformation parameter $\epsilon_4$ we first study an example where we for
clarity exaggerate somewhat the effect.
We show in Fig.~\ref{ndndhugp} the configuration in
Fig.~\ref{ndndgent}, with one change, namely we select $\epsilon_4$ and
$\epsilon_6$ so that a well-developed neck results.
The configuration is [p$^-$,p$^-$,\oriec]. The corresponding
calculated fusion-barrier parameters are listed on line 6 of Table 1.
This hypothetical shape is presented to show the effect
of a well-developed neck on the fusion barrier
and touching configuration. Clearly this configuration is
very different from both the spherical configuration and
the gentle configuration and quite compact. Similar
configurations with necks in the equatorial regions
instead of bulging midsections could favor a large
cross section for complete fusion. Because the nuclei
``grab'' each other we call this configuration corresponding to
this specific orientation and where both
projectile and target exhibit some neck formation {\it
hugging fusion}. In our classification scheme hugging
fusion corresponds to
the [p$^-$,p$^-$,\oriec]
class of touching fusion configurations.
The $\epsilon_4$  deformation value
selected to clearly show this principle is probably unrealistically
large. However, large positive $\epsilon_4$ deformations occur in the
end of the rare-earth region. To compare the effect of a realistic
positive value of $\epsilon_4$  with the effect of a large negative
$\epsilon_4$ we apply the deformation calculated [14] for
$^{186}$W to $^{150}$Nd and display the configuration in
Fig.~\ref{ndndhugn} and the fusion barrier in line 7 of Table 1.
We see that the distance between mass centers of this configuration is
only 10.86 fm, that is, 1.47 fm more compact than the spherical
configuration and 0.88 fm more compact than a configuration with
a large negative $\epsilon_4$. To exploit the enhancement of
the evaporation-residue cross section that we expect in the hugging
configuration [p$^-$,p$^-$,\oriec]
 we must find
suitable projectiles and targets   with large positive $\epsilon_4$
ground-state deformations that lead to superheavy elements with
half-lives that are sufficiently long that the evaporation residues
are observable.

\section{Heavy-ion reactions for distant superheavy-element
  production}

The most stable nuclei on the superheavy island are predicted to occur
in the vicinity of $^{288\mbox{\rm --}294}$110
even though the magic proton number
in this region is calculated to be 114 [22].
However, nuclei at some considerable distance away from
the center of the island are calculated to be sufficiently long-lived
to allow observation after formation; that is, they are predicted to
have half-lives in excess of 1~$\mu$s. We refer to  elements with
proton number larger than 114 as distant superheavy elements.
We now look at some heavy-ion reactions that may lead to
this far part of the superheavy island.

\subsection{Hugging fusion candidates for distant superheavy-element
production}

Above we noted that to achieve very compact configurations of deformed
nuclei one should find projectiles and targets  with large positive
values of the $\epsilon_4$ deformation parameter. Clearly then, the
best candidates for a stable target
above proton number 50 would  be nuclei near the
end of the rare-earth region. To be specific, we select $^{186}$W as a
target in our first example. For this nucleus, calculations [14]
give $\epsilon_4=0.100$ and
$\epsilon_6=-0.044$. The large negative value
of $\epsilon_6$ also
contributes to the development of a neck. A suitable projectile that
would take us to the region of distant superheavy elements would then
be $^{110}$Pd leading to the compound system $^{296}$120. The hugging
configuration for this choice is shown from four different angles in
Fig.~\ref{wpd4}. The fusion barrier for the hugging configuration
[p$^-$,p$^-$,\oriec] is  listed  on line 12 of Table 1,
where we to illustrate the orientation effect on the fusion
barrier also  list the barrier parameters for the four other deformed
configurations
[p$^-$,p$^-$,\oript],
[p$^-$,p$^-$,\oripp],
[p$^-$,p$^-$,\oriet] and
[p$^-$,p$^-$,\oriep] on lines 8--11.
These five deformed configurations also appear in Fig.~\ref{oripmpm}
for slightly different projectile-target sizes and deformations.
The table listing on lines 8--12 is in the order the configurations
occur in
Fig.~\ref{oripmpm}.
In Table 1 we also list on line 13, for reference,
the  barrier parameters for the [s,s,\oris] configuration.

To make an estimate of the decay properties of the compound system we make
the following assumptions.
The heavy-ion reaction takes place at the fusion-barrier energy.
We do not calculate the branching ratio between fusion-fission
and complete fusion, but are primarily interested in studying
the alpha-decay rates of the compound nuclei that possibly do not
fission but de-excite by neutron emission.
One expects of course that at high excitation energy some washing
out of shell effects has taken place and that
$\Gamma_{\rm f}/\Gamma_{\rm n}$ is large. It is a remaining,
important problem to calculate this quantity.
We assume that
neutrons are emitted as long as energetically possible.
The $Q$-values and masses required for these calculations are obtained
from Ref.~[14]. The $\alpha$-decay half-lives are
calculated as discussed in Ref.~[22].
With these assumptions we find for the reaction
and configuration [p$^-$,p$^-$,\oriec]
shown in Fig.~\ref{wpd4} at a center-of-mass energy
equal to  the Coulomb barrier energy listed on line 12 in Table
1 that two neutrons are emitted. Thus
\beq
^{110}{\rm Pd}+ \mbox{}^{186}{\rm W}
\rightarrow \mbox{}^{296}120^*
\rightarrow \mbox{}^{294}120 + {\rm 2n}
\eeq{react1a}
where the compound nucleus has an excitation energy of 35.04 MeV
before neutron emission. The $\alpha$-decay-chain half-lives and
$Q$-values are shown in Fig.~\ref{chainpdw}. Although the first few
decays are calculated to be only a few $\mu$s, these decays should
be within the detection limit of SHIP. Fission half-life calculations
are characterized by large uncertainties [22], but
the calculated ground-state microscopic corrections in the
region of the compound system are about $-7$~MeV, so one expects a
fission barrier about this high in this region of nuclei.
Such a high barrier would probably be associated with fission
half-lives that are longer than the calculated $\alpha$ half-lives
down to about element 104 for all the decay chains considered here.

We have also considered the reaction
\beq
^{104}{\rm Ru}+ \mbox{}^{192}{\rm Os}
\rightarrow \mbox{}^{296}120^*
\rightarrow \mbox{}^{294}120 + {\rm 2n}
\eeq{react1b}
The barrier parameters are listed  on line 14 in Table 1. A
beam energy equal to the Coulomb barrier value of 367.82 MeV leads to
a compound-nucleus excitation energy of 34.06 MeV,
which is about 1 MeV lower
than in the reaction~(\ref{react1a}), and consequently to the
same $\alpha$-decay sequence after 2n emission.

We have also calculated the fusion-barrier parameters for the
[p$^-$,p$^-$,\oriec]
configuration of the reaction
\beq
^{104}{\rm Ru}+ \mbox{}^{186}{\rm W}
\rightarrow \mbox{}^{290}118^*
\rightarrow \mbox{}^{288}118 + {\rm 2n}
\nonumber
\eeq{react1c}
with the result that the compound system $^{290}$118 is created
at an excitation energy of 38.67 MeV for a beam energy at the Coulomb
barrier, leading to 2n emission.
The barrier parameters are listed  on line 15 in Table 1.
The resulting $\alpha$-decay chain is
two neutrons more deficient than the reaction~(\ref{react1a}) and is
shown in Fig.~\ref{chainruwet}.

\subsection{Oblate fusion configurations}

We showed above that the sign of $\epsilon_4$ significantly
influenced fusion-barrier properties of colliding heavy ions.
Correspondingly, one anticipates that the sign of $\epsilon_2$ has a
large effect on the fusion barrier, so in Figs.~\ref{wcdkok} and
\ref{wcdbeko} we give two examples of fusion configurations
corresponding to an oblate projectile and a prolate target for
the reaction
\beq
^{116}{\rm Cd}+ \mbox{}^{186}{\rm W}\rightarrow
\mbox{}^{302- x{\rm n}}122 + {\rm \mbox{$x$}n}
\eeq{reac2}
We have taken the ground-state deformations from the mass calculation
of Ref.~[14]. Experimentally the sign of the ground-state
quadrupole deformation is sometimes not known and may be different
from the calculated value. The examples in this section are
therefore  intended only as examples of how oblate deformations
affect fusion properties.
The
corresponding barrier parameters, as well as the parameters for the
spherical reference case are listed on lines 16--18 in Table 1.
Because the configuration [o,p$^-$,$_|^|$] shown in Fig.~\ref{wcdkok}
resembles
a Japanese Kokeshi doll, although our head here is somewhat large,
we call this configuration as well as the configuration [o,p$^+$,$_|^|$]
Kokeshi.
The
Kokeshi configuration is somewhat more compact than the spherical case
at a cost of only a marginally higher barrier.  The configuration
shown in Fig.~\ref{wcdbeko} is considerably more compact than the
Kokeshi configuration, but at the cost of an 18 MeV higher fusion
barrier. The calculated neutron-emission
and $\alpha$-decay chains of the compound nuclei corresponding to
the initial configurations in Figs.~\ref{wcdkok} and \ref{wcdbeko}
are shown in Figs.~\ref{chaincdwpp} and \ref{chaincdwet},
respectively. For both orientations we have chosen the beam energy
as the Coulomb barrier listed for the configuration in Table 1.

\subsection{An earlier favorite}

Because the reaction
\beq
^{48}{\rm Ca}+ \mbox{}^{248}{\rm Cm}\rightarrow
\mbox{}^{296- x{\rm n}}116 + {\rm \mbox{$x$}n}
\eeq{reac3}
leads to a compound system near the doubly magic $^{298}$114 in
the superheavy region, and because the asymmetric entrance-channel
configuration was thought to enhance the evaporation-residue cross
section it was a favorite reaction in the past. However,
several experiments at different beam energies based on this reaction
did not lead to the identification of any evaporation residues in the
superheavy region [23]. This reaction involves a
spherical
projectile  and a deformed target. Thus, there are two limiting
orientations, polar and equatorial, which are shown in
Figs.~\ref{cmcap} and \ref{cmcae}, respectively, with the
corresponding fusion-barrier parameters listed on lines
19 and 20 of Table 1. From the figures and table it is clear
that the equatorial configuration
$[{\rm s},{\rm p}^+,$\orie $]$
is much more compact than the polar configuration
$[{\rm s},{\rm p}^+,$\orip $]$ and consequently the equatorial
configuration should be the most favorable for heavy
element-production. Although several beam energies  in the range
between the polar and equatorial Coulomb barriers were investigated
in the experiments [23]
there are several possibilities why no heavy elements were observed.

First, the cross-section limit at which detection was possible was
several orders of magnitude higher than the limits that have been
achieved today.

Second, in cold fusion there is appreciable cross section only in
a narrow range, a few MeV, of beam energy [4]. Also in
hot fusion the beam energy may be critically important and it is not
clear that any of the beam energies in the experiment
[23]
corresponded to maximum cross section, in particular since no accurate
theoretical model exists for the evaporation-residue cross section as
a function of beam energy and nuclear orientation. If compact fusion
configurations
are favorable one would in analogy with cold fusion suspect that
one should choose a beam energy slightly below the Coulomb barrier in
the equatorial direction. We show in Fig.~\ref{chaincacme}
the $\alpha$-decay chain that is calculated to occur from the compound
system after neutron emission, when the compound system is formed in
a $[{\rm s},{\rm p}^+,$\orie $]$ configuration at a beam energy
corresponding to the Coulomb barrier in this direction.

Third, the rare cold-fusion events that lead to heavy-element formation
have been described in terms of {\it fusion initiated by transfer}
(FIT) [24].
In this description the occupancy of levels near the Fermi
surface suggests that nucleons will transfer from target to projectile,
resulting in increasing Coulomb repulsion.
In the FIT model one therefore expects
a very low evaporation-residue cross section
for this reaction [24].

\section{Summary}

In heavy-ion collisions between deformed projectiles and targets
we have shown that the fusion reaction
depends strongly on the relative orientation of the
projectile and target.
Both the fusion-barrier height and the compactness of the
touching configuration are so strongly affected that
a variation of relative orientation may have a similar impact as
varying the projectile  and/or target nuclear species. Therefore,
a detailed consideration of deformation is necessary in both theory
and experimental work so that we can understand more about
the many features of heavy-ion reactions between deformed nuclei.
To facilitate such studies we have introduced a classification scheme
of deformed fusion configurations.

Systematic experimental work on understanding cold-fusion reactions
and associated cross sections for evaporation-residue formation
and parallel investigations of microscopic nuclear-structure models
have over the last 20 years or so led to the discovery of five new
elements on the side of the superheavy island closest to us.  Similar
or more extensive work will be required to describe in detail the
fusion reactions between two deformed nuclei. However, the reward may
be access to the far side of the superheavy island.  Of particular
interest is to study how the high charge numbers of these nuclei
affect nuclear and atomic properties. Above we have given a few
examples of heavy-ion reactions that could serve as particularly
suitable starting points for exploring both theoretically and
experimentally the new physics of deformed heavy-ion reactions, and
possibly the new physics of the far side of the superheavy island.
In particular we have suggested that a few special fusion configurations may
be especially favorable for forming superheavy elements. In hot fusion,
we suggest as most favorable an asymmetric projectile-target
combination in the {\it hugging} configuration
$[{\rm p}^-,{\rm p}^-,$\oriec $]$.

This work was supported by the Japan Atomic Energy Research
Institute
and by the U.~S.\ Department of
Energy.

\newpage

\markboth
{\protect\small\it A. Iwamoto, P. M\"{o}ller, J. R. Nix, and H. Sagawa /
Collisions of deformed nuclei}
{\protect\small\it A. Iwamoto, P. M\"{o}ller, J. R. Nix, and H. Sagawa /
Collisions of deformed nuclei}
\begin{small}

\begin{thebibliography}{10}

\bibitem{munzenberg81:a}
G.\ {M\"{u}nzenberg}, S.\ Hofmann, F.\ P.\ He{\ss}berger, W.\ Reisdorf, K.-H.\
  Schmidt, J.~R.~H.\ Schneider, P.\ Armbruster, C.-C.\ Sahm, and B.\ Thuma, Z.\
  Phys.\ {\bf A300} (1981)~7.

\bibitem{munzenberg84:a}
G.\ {M\"{u}nzenberg}, P.\ Armbruster, H.\ Folger, F.\ P. He{\ss}berger, S.\
  Hofmann, J.\ Keller, K.\ Poppensieker, W.\ Reisdorf, K.-H.\ Schmidt, H.\ J.\
  {Sch\"{o}tt}, M.\ E.\ Leino, and R.\ Hingmann, Z.\ Phys.\ {\bf A317} (1984)
  235.

\bibitem{munzenberg82:a}
G.\ {M\"{u}nzenberg}, P.\ Armbruster, F.\ P.\ He{\ss}berger, S.\ Hofmann, K.\
  Poppensieker, W.\ Reisdorf, J.\ R.\ H.\ Schneider, W.\ F.\ W.\ Schneider,
  K.-H.\ Schmidt, C.-C.\ Sahm, and D.\ Vermeulen, Z.\ Phys.\ {\bf A309} (1982)
  89.

\bibitem{hofmann95:a}
S.\ Hofmann, N.\ Ninov, F.\ P.\ He{\ss}berger, P.\ Armbruster, H.\ Folger, G.\
  {M\"{u}nzenberg}, H.\ J.\ Sch{\"{o}}tt, A.\ G.\ Popeko, A.\ V.\ Yeremin, A.\
  N.\ Andreyev, S.\ Saro, R.\ Janik, and M.\ Leino, Z.\ Phys.\ {\bf A350}
  (1995) 277.

\bibitem{hofmann95:b}
S.\ Hofmann, N.\ Ninov, F.\ P.\ He{\ss}berger, P.\ Armbruster, H.\ Folger, G.\
  {M\"{u}nzenberg}, H.\ J.\ Sch{\"{o}}tt, A.\ G.\ Popeko, A.\ V.\ Yeremin, A.\
  N.\ Andreyev, S.\ Saro, R.\ Janik, and M.\ Leino, Z.\ Phys.\ {\bf A350}
  (1995) 281.

\bibitem{nix74:a}
J.\ R.\ Nix and A.\ J.\ Sierk, Phys.\ Scr.\ {\bf 10A} (1974) 94.

\bibitem{nix77:a}
J.\ R.\ Nix and A.\ J.\ Sierk, Phys.\ Rev.\ {\bf C15} (1977) 2072.

\bibitem{blocki78:a}
J.\ B{\l}ocki, Y.\ Boneh, J.\ R.\ Nix, J.\ Randrup, M.\ Robel, A.\ J.\ Sierk,
  and W.\ J.\ Swiatecki, Ann.\ Phys.\ (N.\ Y.) {\bf 113} (1978) 330.

\bibitem{davies83:a}
K.\ T.\ R.\ Davies, A.\ J.\ Sierk, and J.\ R.\ Nix, Phys.\ Rev.\ {\bf C28}
  (1983) 679.

\bibitem{moller76:a}
P.\ {M\"{o}ller} and J.\ R.\ Nix, Nucl.\ Phys.\ {\bf A272} (1976) 502.

\bibitem{nilsson55:a}
S.\ G.\ Nilsson, Kgl.\ Danske Videnskab.\ Selskab.\ Mat.-Fys.\ Medd.\ {\bf
  29}:No.\ 16 (1955).

\bibitem{moller94:c}
P.\ M{\"{o}}ller and A.\ Iwamoto, Nucl.\ Phys.\ {\bf A575} (1994) 381.

\bibitem{iwamoto95:a}
A.\ Iwamoto and P.\ M{\"{o}}ller, to be published (1995).

\bibitem{moller95:b}
P.\ M{\"{o}}ller, J.\ R.\ Nix, W.\ D.\ Myers, and W.\ J.\ Swiatecki, {Atomic
  Data Nucl.\ Data Tables} {\bf 59} (1995) 185.

\bibitem{stefanini95:a}
A.\ M.\ Stefanini, D.\ Ackermann, L.\ Corradi, D.\ R.\ Napoli, C.\ Petrache,
  P.\ Spolare, P.\ Bednarczyk, H.\ Q.\ Zhang, S.\ Beghini, G.\ Montagnoli, L.\
  Mueller, F.\ Scarlassara, G.\ F.\ Segato, F.\ Soramel, and N.\ Rowley, Phys.
  Rev. Lett. {\bf 74} (1995) 864.

\bibitem{leigh93:a}
R.\ J.\ Leigh, N.\ Rowley, R.\ C.\ Lemmon, D.\ J.\ Hinde, J.\ O.\ Newton, J.\
  X.\ Wei, J.\ C.\ Mein, C.\ R.\ Morton, S.\ Kuyucak, and A.\ T.\ Kruppa,
  Phys.\ Rev.\ {\bf C47} (1993) 47.

\bibitem{moller81:a}
P.\ {M\"{o}ller} and J.\ R.\ Nix, Nucl.\ Phys.\ {\bf A361} (1981) 117.

\bibitem{moller81:b}
P.\ {M\"{o}ller} and J.\ R.\ Nix, {Atomic Data Nucl.\ Data Tables} {\bf 26}
  (1981) 165.

\bibitem{lemmon93:a}
R.\ C.\ Lemmon, R.\ J.\ Leigh, J.\ X.\ Wei, C.\ R.\ Morton, D.\ J.\ Hinde, J.\
  O.\ Newton, J.\ C.\ Mein, M.\ Dasgupta, and N.\ Rowley, Phys.\ Lett.\ {\bf
  B316} (1993) 32.

\bibitem{norenberg94:a}
W. N{\"{o}}renberg, Proc.\ Int.\ Workshop on Heavy-Ion Fusion, Padua, Italy
  (1994).

\bibitem{norenberg94:b}
W. N{\"{o}}renberg, GSI Nachrichten 10-94 (Oct. 1994) p.~13.

\bibitem{moller94:b}
P.\ M{\"{o}}ller and J.\ R.\ Nix, J.\ Phys.\ G: Nucl.\ Part.\ Phys.\ {\bf 20}
  (1994) 1681.

\bibitem{armbruster85:b}
P.\ Armbruster, Y.\ K.\ Agarwal, W.\ Br{\"{u}}chle, M.\ Br{\"{u}}gger, J.\ P.\
  Dufour, H.\ G{\"{a}}ggeler, F.\ P.\ He{\ss}berger, S.\ Hofmann, P.\ Lemmertz,
  G.\ {M\"{u}nzenberg}, K.\ Poppensieker, W.\ Reisdorf, M.\ Sch{\"{a}}del,
  K.-H.\ Schmidt, J.\ R.\ H.\ Schneider, W.\ F.\ W.\ Schneider, K.\
  S{\"{u}}mmerer, D.\ Vermeulen, G.\ Wirth, A.\ Ghiorso, K.\ E.\ Gregorich, D.\
  Lee, M.\ E.\ Leino, K.\ J.\ Moody, G.\ T.\ Seaborg, R.\ B.\ Welch, P.\
  Wilmarth, S.\ Yashita, C.\ Frink, N.\ Greulich, G.\ Herrmann, U.\ Hickmann,
  N.\ Hildebrand, J.\ V.\ Kratz, N.\ Trautmann, M.\ M.\ Fowler, D.\ C.\
  Hoffman, W.\ R.\ Daniels, H.\ R.\ von Gunten, and H.\ Dornh{\"{o}}fer, Phys.\
  Rev.\ Lett.\ {\bf 54} (1985) 406.

\bibitem{hofmann95:c}
S.\ Hofmann, Proc. XV Nuclear Physics Divisional Conf. on `Low Energy Nuclear
  Dynamics,' St. Petersburg, Russia (1995).

\bibitem{stokstad80:a}
R.\ G.\ Stokstad, Y.\ Eisen, S.\ Kaplanis, D.\ Pelte, U.\ Smilansky, and I.\
  Tserruya, Phys.\ Rev. {\bf C21} (1980) 2427.

\end{thebibliography}

\end{small}
\markboth
{\protect\small\it A. Iwamoto, P. M\"{o}ller, J. R. Nix, and H. Sagawa /
Collisions of deformed nuclei}
{\protect\small\it A. Iwamoto, P. M\"{o}ller, J. R. Nix, and H. Sagawa /
Collisions of deformed nuclei}
\pagebreak
\begin{center}
{\bf Figure Captions}
\end{center}
\newcounter{bean}
\begin{list}
{\phantom{\Roman{bean}}}{\usecounter{bean}
\setlength{\leftmargin}{0.7in}
\setlength{\rightmargin}{0.0in}
\setlength{\labelwidth}{0.60in}
\setlength{\labelsep}{0.10in}
}
\item[Fig.\ \ref{orisphe}\hfill]
The seven limiting touching configurations with
spherical projectiles.
The simplest configuration with a spherical target is in
the top row third from the left.
To the left of this configuration are configurations with prolate
target shapes
whereas  to the right are the two limiting configurations that occur
for oblate target shape.
The ratio between the projectile and target volume is 0.343.
The deformation is
$\beta_2=0.30$ and $\beta_4=0.11$
for p$^+$,
$\beta_2=0.24$ and $\beta_4=-0.09$ for p$^-$, and
$\beta_2=-0.25$ and $\beta_4=0.0$
for o shapes.
The arrows give the direction of the incident beam.
The nuclear symmetry axis is indicated by a thin line emerging from
the nuclear polar regions.

\item[Fig.\ \ref{oripppp}\hfill]
Five limiting touching configurations with
prolate, positive-hexadecapole projectiles
and targets.
Specifically
$\beta_2=0.30$ and $\beta_4=0.11$.
The ratio between the projectile and target volume is 0.343.
Only the relative positions and orientations change between the
configurations. The arrows give the direction of the incident beam.
The nuclear symmetry axis is indicated by a thin line emerging from
the nuclear polar regions.

\item[Fig.\ \ref{oripppm}\hfill]
Five limiting touching configurations with  a
prolate, positive-hexadecapole projectile
and a prolate, negative-hexadecapole target.
Specifically, for the projectile
$\beta_2=0.30$ and $\beta_4=0.11$
and for the target
$\beta_2=0.24$ and $\beta_4=-0.09$.
The ratio between the projectile and target volume is 0.343.
Only the relative positions and orientations change between the
configurations. The arrows give the direction of the incident beam.
The nuclear symmetry axis is indicated by a thin line emerging from
the nuclear polar regions.

\item[Fig.\ \ref{orippo}\hfill]
Five limiting touching configurations with  a prolate,
positive-hexadecapole projectile
and an oblate target.
Specifically, for the projectile
$\beta_2=0.30$ and $\beta_4=0.11$
and for the target
$\beta_2=-0.25$ and $\beta_4=0$.
The ratio between the projectile and target volume is 0.343.
Only the relative positions and orientations change between the
configurations. The arrows give the direction of the incident beam.
The nuclear symmetry axis is indicated by a thin line emerging from
the nuclear polar regions.

\item[Fig.\ \ref{oripmpp}\hfill]
Five limiting touching configurations with  a
prolate, negative-hexadecapole projectile
and a prolate, positive-hexadecapole target.
Specifically, for the projectile
$\beta_2=0.24$ and $\beta_4=-0.09$
and for the target
$\beta_2=0.30$ and $\beta_4=0.11$.
The ratio between the projectile and target volume is 0.343.
Only the relative positions and orientations change between the
configurations. The arrows give the direction of the incident beam.
The nuclear symmetry axis is indicated by a thin line emerging from
the nuclear polar regions.

\item[Fig.\ \ref{oripmpm}\hfill]
Five limiting touching configurations with
prolate, negative-hexadecapole projectiles
and targets.
Specifically
$\beta_2=0.24$ and $\beta_4=-0.09$.
The ratio between the projectile and target volume is 0.343.
Only the relative positions and orientations change between the
configurations. The arrows give the direction of the incident beam.
The nuclear symmetry axis is indicated by a thin line emerging from
the nuclear polar regions.

\item[Fig.\ \ref{oripmo}\hfill]
Five limiting touching configurations with  a prolate,
negative-hexadecapole projectile
and an oblate target.
Specifically, for the projectile
$\beta_2=0.24$ and $\beta_4=-0.09$
and for the target
$\beta_2=-0.25$ and $\beta_4=0$.
The ratio between the projectile and target volume is 0.343.
Only the relative positions and orientations change between the
configurations. The arrows give the direction of the incident beam.
The nuclear symmetry axis is indicated by a thin line emerging from
the nuclear polar regions.

\item[Fig.\ \ref{oriopp}\hfill]
Five limiting touching configurations with  an
oblate  projectile
and a prolate, positive-hexadecapole target.
Specifically, for the projectile
$\beta_2=-0.25$ and $\beta_4=0.0$
and for the target
$\beta_2=0.30$ and $\beta_4=0.11$.
The ratio between the projectile and target volume is 0.343.
Only the relative positions and orientations change between the
configurations. The arrows give the direction of the incident beam.
The nuclear symmetry axis is indicated by a thin line emerging from
the nuclear polar regions.

\item[Fig.\ \ref{oriopm}\hfill]
Five limiting touching configurations with  an
oblate  projectile
and a prolate, negative-hexadecapole target.
Specifically, for the projectile
$\beta_2=-0.25$ and $\beta_4=0.0$
and for the target
$\beta_2=0.24$ and $\beta_4=-0.09$.
The ratio between the projectile and target volume is 0.343.
Only the relative positions and orientations change between the
configurations. The arrows give the direction of the incident beam.
The nuclear symmetry axis is indicated by a thin line emerging from
the nuclear polar regions.

\item[Fig.\ \ref{orioo}\hfill]
Five limiting touching configurations with
oblate projectiles and  targets.
Specifically,
$\beta_2=-0.25$ and $\beta_4=0$.
The ratio between the projectile and target volume is 0.343.
Only the relative positions and orientations change between the
configurations. The arrows give the direction of the incident beam.
The nuclear symmetry axis is indicated by a thin line emerging from
the nuclear polar regions.

\item[Fig.\  \ref{potsmod}\hfill]
Calculated
potential-energy surface in units of MeV
for the reaction $^{16}{\rm O} \mbox{}+\mbox{}^{154}$Sm.
The energy in the medium-gray area, outside the
dark-grey $^{154}$Sm nucleus in the
center was not calculated, because the points in this region correspond
to points inside the touching configuration.  The light-gray
shape inside the
$^{154}$Sm nucleus has been drawn to show the relative size of the
projectile to the target. This overlapping configuration is not
considered in actual calculations.  Note the ridge with saddle points  and
peaks around the target nucleus.

\item[Fig.\  \ref{crsmo}\hfill]
Calculated fusion cross sections for the reaction
$^{16}{\rm O}  \mbox{}+ \mbox{} ^{154}$Sm,
compared to experimental data~[25].
The solid curve corresponds to the fusion surface shown in
Fig.~\ref{potsmod} when the shape of the target corresponds
to the calculated ground-state shape. The long-dashed curve is the cross
section obtained for a hypothetical spherical  target.
The arrows show the fusion-barrier height in the polar direction
(p), the equatorial plane (e), and the barrier height for a
hypothetical
spherical target (s).
Both the curves and the arrows have
been translated in energy by $E_{\rm tran} = - 3.1 $ MeV from
their calculated values.

\item[Fig.\  \ref{ndndgent}\hfill]
Touching configuration of
$^{150}{\rm Nd} \mbox{}+\mbox{}^{150}$Nd  with the nuclear
shapes taken to be the calculated [14] ground-state
shape;
that is,  the configuration is $[{\rm p}^+,{\rm p}^+,$\oriec $]$.
The arrow gives the direction of the incident beam.
Fusion-barrier parameters for this configuration/direction
are given on line 5 of Table 1.

\item[Fig.\  \ref{ndndhugp}\hfill]
Touching configuration of $^{150}{\rm Nd}+\mbox{}^{150}$Nd for
hypothetical nuclear shapes with a large positive $\epsilon_4$
and a choice of $\epsilon_6$ that further develops the waistline;
that is,  the configuration is $[{\rm p}^-,{\rm p}^-,$\oriec $]$.
The arrow gives the direction of the incident beam.
Fusion-barrier parameters for this configuration/direction
are given on line 6 of Table 1.

\item[Fig.\  \ref{ndndhugn}\hfill]
Touching configuration of $^{150}{\rm Nd} \mbox{}+\mbox{}^{150}$Nd for
shapes corresponding to a realistic ground-state shape in
the rare earth region;
that is,  the configuration is $[{\rm p}^-,{\rm p}^-,$\oriec $]$.
Specifically, we have chosen the ground-state
shape parameters of $^{186}$W.
Fusion-barrier parameters for this configuration/direction
are given on line 7 of Table 1.

\item[Fig.\  \ref{wpd4}\hfill]
Touching configuration of $^{110}{\rm Pd} \mbox{}+\mbox{}^{186}$W for
calculated ground-state shapes viewed from four different angles.
The shapes used are the calculated ground-states shapes, so the
configuration is $[{\rm p}^-,{\rm p}^-,$\oriec $]$.
The arrows and $\bigotimes$ sign give the direction of the incident beam.
Fusion-barrier parameters for this configuration/direction
are given on line 12 of Table 1.

\item[Fig.\  \ref{chainpdw}\hfill]
Calculated $Q$-values for $\alpha$ decay and corresponding calculated
half-lives for the decay chain starting
at $^{294}120$.

\item[Fig.\  \ref{chainruwet}\hfill]
Calculated $Q$-values for $\alpha$ decay and corresponding calculated
half-lives for the decay chain starting
at $^{288}118$.

\item[Fig.\  \ref{wcdkok}\hfill]
Polar-parallel Kokeshi configuration of $^{116}{\rm Cd}+\mbox{}^{186}$W for
shapes corresponding to calculated ground-state  deformation parameters;
that is,  the configuration is $[{\rm o},{\rm p}^-,$\oripp $]$.
The arrow gives the direction of the incident beam.
Fusion-barrier parameters for this configuration/direction
are given on line 16 of Table 1.

\item[Fig.\  \ref{wcdbeko}\hfill]
Equatorial-transverse configuration of $^{116}{\rm Cd}+\mbox{}^{186}$W for
shapes corresponding to calculated ground-state deformation parameters;
that is,  the configuration is $[{\rm o},{\rm p}^-,$\oriet $]$.
The arrow gives the direction of the incident beam.
Fusion-barrier parameters for this configuration/direction
are given on line 17 of Table 1.

\item[Fig.\  \ref{chaincdwpp}\hfill]
Calculated $Q$-values for $\alpha$ decay and corresponding calculated
half-lives for the decay chain starting
at $^{301}122$.

\item[Fig.\  \ref{chaincdwet}\hfill]
Calculated $Q$-values for $\alpha$ decay and corresponding calculated
half-lives for the decay chain starting
at $^{300}122$.

\item[Fig.\  \ref{cmcap}\hfill]
Touching configuration of $^{48}{\rm Ca} \mbox{}+\mbox{}^{248}$Cm for
shapes corresponding to calculated ground-state deformation parameters.
The arrow gives the direction of the incident beam.
Fusion-barrier parameters for this configuration/direction
are given on line 19 of Table~1.

\item[Fig.\  \ref{cmcae}\hfill]
Touching configuration of $^{48}{\rm Ca} \mbox{}+\mbox{}^{248}$Cm for
shapes corresponding to calculated ground-state deformation
parameters.
The arrow gives the direction of the incident beam.
Fusion-barrier parameters for this configuration/direction
are given on line 20 of Table~1.

\item[Fig.\  \ref{chaincacme}\hfill]
Calculated $Q$-values for $\alpha$ decay and corresponding calculated
half-lives for the decay chain starting
at $^{293}116$.

\end{list}
\newpage
\begin{enumerate}
\item
   \label{orisphe}
\item
   \label{oripppp}
\item
   \label{oripppm}
\item
   \label{orippo}
\item
   \label{oripmpp}
\item
   \label{oripmpm}
\item
   \label{oripmo}
\item
   \label{oriopp}
\item
   \label{oriopm}
\item
   \label{orioo}
\item
   \label{potsmod}
\item
   \label{crsmo}
\item
   \label{ndndgent}
\item
   \label{ndndhugp}
\item
   \label{ndndhugn}
\item
   \label{wpd4}
\item
   \label{chainpdw}
\item
   \label{chainruwet}
\item
   \label{wcdkok}
\item
   \label{wcdbeko}
\item
   \label{chaincdwpp}
\item
   \label{chaincdwet}
\item
   \label{cmcap}
\item
   \label{cmcae}
\item
   \label{chaincacme}
\end{enumerate}
\end{document}